\documentclass[letterpaper,11pt]{article}
\usepackage{amsmath}
\usepackage{amssymb}
\usepackage[dvips]{graphicx}
\begin{document}
\title{Analytical Potential-Density Pairs for Flat  Rings and Toroidal Structures}
\author{D. Vogt\thanks{e-mail: dvogt@ime.unicamp.br} 
\and
P. S. Letelier\thanks{e-mail: letelier@ime.unicamp.br}\\
Departamento de Matem\'{a}tica Aplicada-IMECC, Universidade \\ 
Estadual de Campinas 13083-970 Campinas, S.\ P., Brazil}
\maketitle
\begin{abstract}
The Kuzmin-Toomre family of discs is used to construct potential-density pairs that
represent flat ring structures in terms of elementary functions. Systems composed of 
two concentric flat rings, a central disc surrounded by one ring and a ring with a centre 
of attraction are also presented. The circular velocity of test particles and the epicyclic 
frequency of small oscillations about circular orbits are calculated for these structures. 
A few examples of three-dimensional
potential-density pairs of ``inflated'' flat rings (toroidal mass distributions) are presented.

\end{abstract}
\section{Introduction}

The presence of flat ring structures is a characteristic feature of the four giant
planets of the Solar System, in particular, the famous Saturn's rings. 
Planetary rings are the thinnest of all astrophysical discs, with a ratio of the
thickness to the radius of order $10^{-6}$ \cite{f1}. On a
larger scale, several ring galaxies, or R galaxies, objects in the form of
 approximate elliptical
rings with no luminous matter visible in their interiors, are known 
(see, for instance, \cite{t1,t2}).
Sometimes, as the result of interactions between galaxies, a ring of gas and stars is formed
and rotates over the poles of a galaxy, originating the so called polar-ring galaxies
(\cite{w1} and references therein).
The exact gravitational potential of a ring of zero thickness and constant linear
density is given by an elliptic integral, which in practice is often approximated by a truncated 
multipolar expansion.

The potential of a disc with negative mass density whose rim has positive mass density 
can be obtained by a process of complexification of the potential of a punctual mass 
\cite{a1,w2,g1}. This potential was used by Letelier and Oliveira \cite{l3} to construct superpositions of
static bodies with axial symmetry in the context of General Relativity. An exact solution
representing the superposition of a disc with a central hole and a Schwarzschild 
black hole was obtained by Lemos and Letelier \cite{l1}. The properties of this space-time and
of space-times that represent superpositions of a Schwarzschild black-hole and a ring
were investigated in \cite{c2,s2,s3}.

Recently,  Letelier \cite{l2} constructed several families of flat rings using the superposition
of Morgan and Morgan \cite{m2} discs of different densities. The main advantage of these models is that
the density and the gravitational potential are given in terms of elementary functions.
The main purpose of this work is to use the same method to construct other potential-density pairs
representing flat ring structures by superposing members of the classical Kuzmin-Toomre family of
discs. Because the Kuzmin-Toomre discs have infinite extension, so do the resulting rings,
while the Letelier rings have a finite outer radius or a central hole. All the potential-density 
pairs can be explicitly stated in terms of elementary functions. We construct
systems composed of one and two concentric flat rings and a central Kuzmin-Toomre 
disc surrounded by a flat ring. By using a 
suitable transformation, some of these flat ring systems are used to generate 
three-dimensional potential-density pairs with toroidal mass distributions. Similar toroidal 
analytical systems have been recently obtained by Ciotti and Giampieri \cite{c4}, and 
Ciotti and Marinacci \cite{c5} using a different technique, 
known as complex-shift method. 

The article is divided as follows. In Section \ref{sec_disc},  we briefly revise  the 
potential-density pair of the Kuzmin-Toomre family of discs. In Section \ref{sec_flat}, 
the discs are superposed to generate potential-density pairs that represent structures
of one and two concentric flat rings, and a central disc surrounded by one ring. We
calculate the circular velocity of particles moving inside these structures and study the
stability under radial perturbations of circular orbits on the plane of the rings. We 
also briefly discuss  the influence of a 
point mass located at the centre of a ring on the rotation profile and on the stability 
of the ring. In Section \ref{sec_thick},  we ``inflate'' some flat ring models by using 
a transformation proposed by Miyamoto and Nagai \cite{m1}. The resulting three-dimensional 
potential-density pairs have toroidal concentrations of matter. In Section \ref{sec_discuss}, we 
summarize and discuss the results. In Appendices \ref{ap_A} to \ref{ap_F}, we collect, 
for reference, several equations expressing potential-density pairs, circular velocities 
and epicyclic frequencies for the structures discussed in the main text. 
\section{Kuzmin-Toomre Discs} \label{sec_disc}

Kuzmin \cite{k1} derived a very simple potential-density pair of an axisymmetric disc. The
potential in cylindrical coordinates $(R,z,\varphi)$ is given by
\begin{equation} \label{eq_phi_kuz}
\phi_0=-\frac{GM}{\sqrt{R^2+(a+|z|)^2}} \mbox{.}
\end{equation}
By Poisson's equation, the discontinuous normal derivative on $z=0$ introduces a
surface density of mass
\begin{equation} \label{eq_sigma_kuz}
\sigma(R)=\frac{1}{2\pi G}\left. \frac{\partial \phi}{\partial z}\right|_{z=0}=
\frac{aM}{2\pi \left( R^2+a^2 \right)^{3/2}} \mbox{.}
\end{equation}
The potential (\ref{eq_phi_kuz}) is in fact the first member of the Kuzmin-Toomre
family of discs \cite{t3}.  The other members can be calculated by using the following
recurrence relations \cite{n1,b1}
\begin{gather}
\phi_{n+1} =\phi_{n}-\frac{a}{2n+1} \frac{\partial}{\partial a} \phi_{n} \mbox{,} \\
\sigma_{n+1} =\sigma_{n}-\frac{a}{2n+1} \frac{\partial}{\partial a} \sigma_{n}
\text{,} \qquad n=0,1,\ldots \mbox{,}
\end{gather}
where $\phi_0$, $\sigma_0$ is the Kuzmin potential-density pair
(\ref{eq_phi_kuz})--(\ref{eq_sigma_kuz}). The general expression for the surface 
density of the \emph{n}th-order Kuzmin-Toomre disc is \cite{n1}
\begin{equation} \label{eq_sigma_gen}
\sigma_{n}=\frac{(2n+1)a^{2n+1}M}{2\pi (R^2+a^2)^{n+3/2}} \mbox{.}
\end{equation}
For reference the density and potential for the members with $n=1,2,3$ are listed
in Appendix \ref{ap_A}.

\section{Flat Rings} \label{sec_flat}
\subsection{A Family of Single Rings} \label{ss_one}

In order to generate ring structures we take Kuzmin-Toomre discs with mass 
\begin{equation}
M=\frac{2\pi a^2\sigma_c}{2n+1} \mbox{,}
\end{equation}
where $\sigma_c$ is a constant with dimensions of surface density. 
Thus the surface density (\ref{eq_sigma_gen}) is
rewritten as 
\begin{equation} \label{eq_sigma_ad}
\sigma_{n}=\frac{\sigma_ca^{2n+3}}{(R^2+a^2)^{n+3/2}}=
\frac{\sigma_c}{\left(1+R^2/a^2 \right)^{n+3/2}}\mbox{.}
\end{equation}
Now let us consider the following superposition
\begin{align} \label{eq_sigma_sum}
\sigma^{(m,n)} &=\sum_{k=0}^m C^m_k(-1)^{m-k}\sigma_{n+m-k} \\
&=\frac{\sigma_c}{\left(1+R^2/a^2 \right)^{n+m+3/2}}\sum_{k=0}^m C^m_k(-1)^{m-k}
\left(1+\frac{R^2}{a^2} \right)^k \mbox{,}
\end{align}
where $C^m_k=m!/[(m-k)!k!]$. Using
\begin{equation}
\sum_{k=0}^m C^m_k(-1)^{m-k}\left(1+\frac{R^2}{a^2} \right)^k=\left( \frac{R}{a} \right)^{2m} \mbox{,}
\end{equation}
equation (\ref{eq_sigma_sum}) takes the form
\begin{equation} \label{eq_sigma_r1}
\sigma^{(m,n)}=\frac{\sigma_c\left( R/a \right)^{2m}}{\left(1+R^2/a^2 \right)^{n+m+3/2}} \mbox{.}
\end{equation}
We have that (\ref{eq_sigma_r1}) with $m=1,2,\ldots$; $n=0,1,\ldots$, defines a family
of flat rings (characterized by zero density on $R=0$) with infinite extension. For large 
$R$, the density decays as $1/R^{2n+3}$; thus in principle, a cut-off radius may be defined
and this flat ring is considered as finite. The density has a maximum at
$\frac{R}{a}=\bigl(\frac{m}{n+3/2} \bigr)^{1/2}$. The total mass $\mathcal{M}^{(m,n)}$ 
of member $(m,n)$ is given by
\begin{equation} \label{eq_mass1}
\mathcal{M}^{(m,n)}=\int_{R=0}^{\infty} \, \int_{\varphi=0}^{2\pi} \sigma^{(m,n)}R \, \mathrm{d}R 
\, \mathrm{d}\varphi=\frac{\Gamma(m+1)\Gamma(n+1/2)}{\Gamma(m+n+3/2)}\pi \sigma_ca^2 \mbox{,}
\end{equation} 
where $\Gamma(x)$ is the gamma function. As $m,n$ are integers, (\ref{eq_mass1}) can be
further simplified to
\begin{equation} \label{eq_mass2}
\mathcal{M}^{(m,n)}=\frac{2^{m+1}m!(2n-1)!!}{(2m+2n+1)!!}\pi\sigma_ca^2 \mbox{.}
\end{equation}

The potentials associated with
(\ref{eq_sigma_r1}) can be calculated using a superposition with the same coefficients 
as in (\ref{eq_sigma_sum}), e.g., the potential associated with $\sigma^{(1,0)}$ is 
$\phi^{(1,0)}=\phi_{0}-\phi_{1}$, etc. The explicit potential-density pairs of four members
with $m=1,2$; $n=0,1$ are given in Appendix \ref{ap_B}.

\begin{figure}
\centering
\includegraphics[scale=0.85]{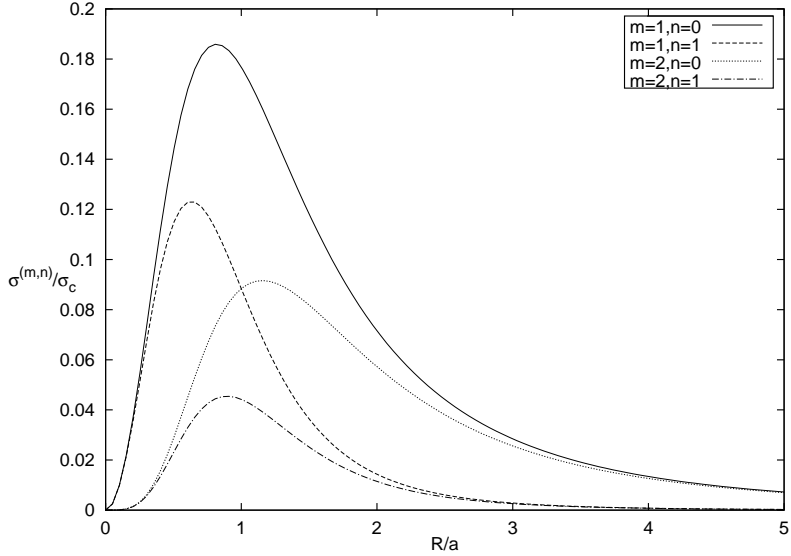}
\caption{The surface density $\sigma^{(m,n)}/\sigma_c$ equation (\ref{eq_sigma_r1}) 
of four members with $m=1,2$; $n=0,1$ of the family of flat single rings as function of $R/a$.} \label{fig_r1}
\end{figure}

In Fig.\ \ref{fig_r1} we show the curves of $\sigma^{(m,n)}/\sigma_c$ as function of $R/a$ for
the four members of rings listed in equations (\ref{eq_r10})--(\ref{eq_r21}). The rings with
$m=2$ have larger central holes than those with $m=1$. Also note how fast the
density decays for $n=1$ $(\propto R^{-5})$. These four members have masses
$\mathcal{M}^{(1,0)}=4\pi\sigma_ca^2/3$, $\mathcal{M}^{(1,1)}=4\pi\sigma_ca^2/15$, 
$\mathcal{M}^{(2,0)}=16\pi\sigma_ca^2/15$ and $\mathcal{M}^{(2,1)}=16\pi\sigma_ca^2/105$.   

Some other physical quantities of interest are the circular velocity $v_c$ of test 
particles concentric to the flat rings and the epicyclic frequency $\kappa$ of small 
oscillations about the equilibrium circular orbit in the plane of the ring.
They are calculated with the expressions \cite{b2}
\begin{gather}
v_c^2=R\phi_{,R} \mbox{,} \label{eq_vc_gen} \\
\kappa^2=\phi_{,RR}+\frac{3\phi_{,R}}{R} \mbox{,} \label{eq_k_gen} 
\end{gather}
where the subscripts with comma indicate partial derivatives and 
all expressions are evaluated on $z=0$. The criterion for stability is given by
$\kappa \geq 0$. This study of stability can be considered as an order 
zero test of stability, where the collective behaviour of the particles of the ring is 
not taken into account.

Using the potential (\ref{eq_r10}), we obtain the following expression for the rotation 
curve of the first member of the family of single rings
\begin{equation}
v_c^{(1,0)} =R \left[ \frac{2\pi \sigma_cGa^2(2R^2-a^2)}{3(R^2+a^2)^{5/2}} \right]^{1/2} \mbox{,}
\end{equation}
We have that circular orbits are only possible in the region 
$R/a \geq \sqrt{2}/2 \approx 0.71$. Near the hole of the ring, there is 
not enough mass to produce the necessary gravitational force to keep a
particle in a circular orbit, however  this can be compensated if a massive point mass 
is placed at the centre of the ring (see Section \ref{ss_rpl}). The rotation curves and the expressions 
for the epicyclic frequency $\kappa^{(m,n)}$ of the other members are listed in Appendix \ref{ap_C}.

\begin{figure}
\centering
\includegraphics[scale=0.85]{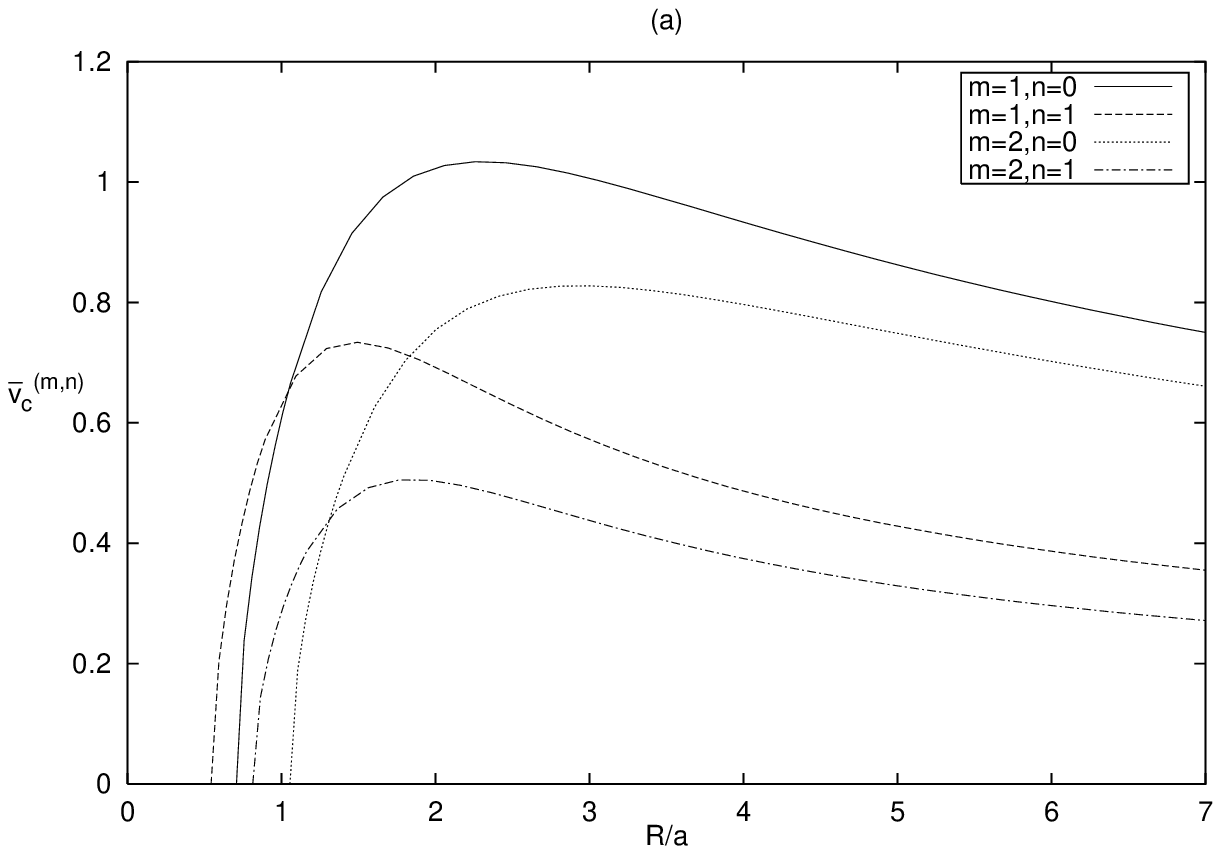}  \\
\vspace{0.1cm}
\includegraphics[scale=0.85]{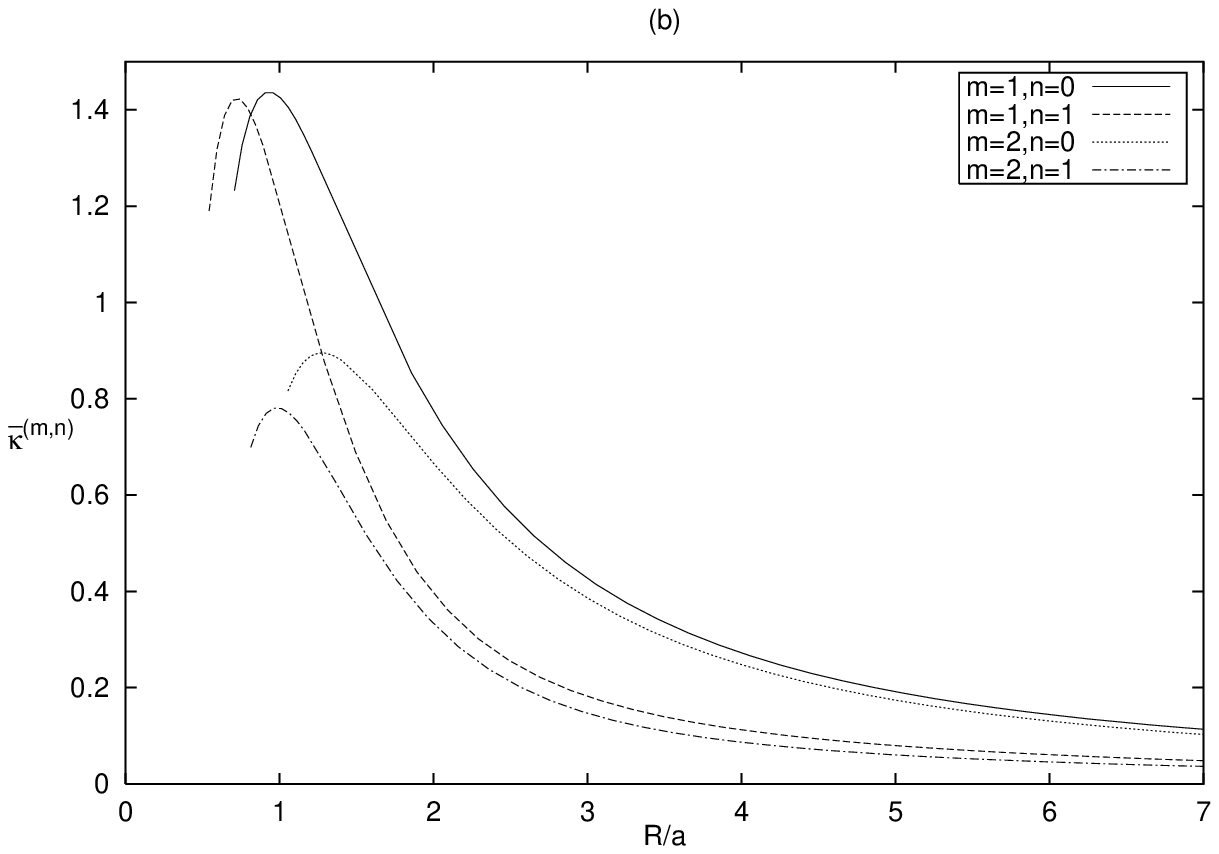}
\caption{(a) The circular velocity $\bar{v}_c^{(m,n)}=v_c^{(m,n)}/\sqrt{\sigma_cGa}$ 
equations (\ref{eq_vc10})--(\ref{eq_vc21}) and (b) the epicyclic frequency 
$\bar{\kappa}^{(m,n)}=\kappa^{(m,n)}/(\sigma_cG/a)^{1/2}$ equations
(\ref{eq_kappa10})--(\ref{eq_kappa21}) as functions of $R/a$ for the four members with
$m=1,2$; $n=0,1$ of the family of flat single rings.} \label{fig_r2}
\end{figure}
In Fig.\ \ref{fig_r2}(a)--(b) we present, respectively, the curves 
of the circular velocity $\bar{v}_c^{(m,n)}=v_c^{(m,n)}/\sqrt{\sigma_cGa}$ and of the
epicyclic frequency $\bar{\kappa}^{(m,n)}=\kappa^{(m,n)}/(\sigma_cG/a)^{1/2}$ as functions
of $R/a$ for our four members of rings. We note that the rings with $m=2$ have larger
regions where circular orbits are not possible than rings with $m=1$ because the former 
have also larger central holes. The possible circular orbits shown are all
stable under radial perturbations. In Fig.\ \ref{fig_r2}(b) we only plotted the
curves of epicyclic frequency in the ranges where circular orbits are possible. 
The intervals of stability shown in equations (\ref{eq_kappa10})--(\ref{eq_kappa21}) 
suggest that rings with larger central holes $(m=2)$ also have more unstable orbits.  
\subsection{A Family of Double Rings} \label{ss_two}

Now we consider the following superposition of flat rings,
\begin{equation} \label{eq_sigma_r2}
\sigma^{(m,n)}_{(2)}=\sigma^{(m,n)}-2k^2\sigma^{(m+1,n-1)}+k^4\sigma^{(m+2,n-2)}=
\frac{\sigma_c\left( R/a \right)^{2m}\left( 1-k^2R^2/a^2\right)^2}{\left(1+R^2/a^2 \right)^{n+m+3/2}} \mbox{,}
\end{equation}
where $k$ is a constant. With this superposition, a gap is placed at $R/a=1/k$ on 
a single ring, thus generating a family of two concentric flat rings. Using equation (\ref{eq_mass2}), the total mass of this 
superposition is found to be
\begin{multline} \label{eq_mass_r2} 
\mathcal{M}^{(m,n)}_{(2)}=\frac{\pi\sigma_ca^22^{m+1}m!(2n-5)!!}{(2m+2n+1)!!} \bigl[
(2n-1)(2n-3)-4k^2(m+1)(2n-3) \bigr. \\
\bigl. +4k^4(m+1)(m+2) \bigr] \mbox{.}
\end{multline}

\begin{figure}
\centering
\includegraphics[scale=0.8]{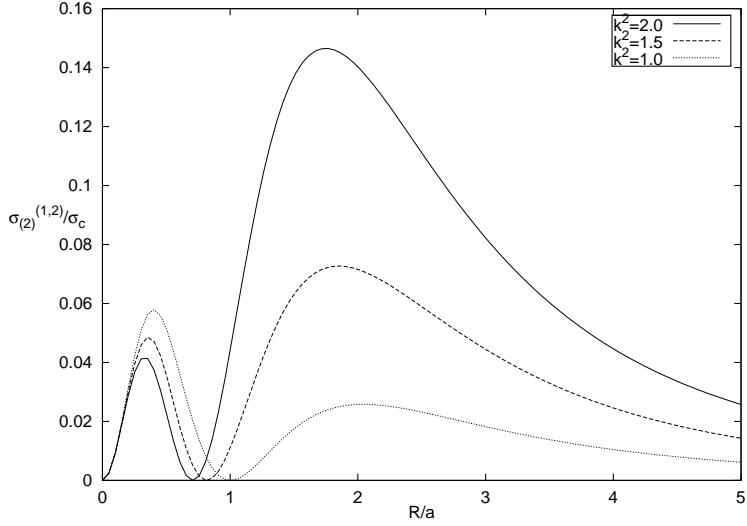}
\caption{The surface density $\sigma^{(1,2)}_{(2)}/\sigma_c$ of the first member of
a family of double rings equation (\ref{eq_sigma_r2}), as function of $R/a$, for $k^2=2$, $k^2=1.5$ and $k^2=1$.} \label{fig_r4}
\end{figure}

We study in detail the first member of this family with $m=1$, $n=2$. Its non-dimensional
surface density is shown in Fig.\ \ref{fig_r4} for the values $k^2=2$, $k^2=1.5$ and 
$k^2=1$. The size of the gap increases for smaller $k$.

The associated potential $\phi^{(1,2)}_{(2)}$ is given by
\begin{multline} \label{eq_phi_r2}
\phi^{(1,2)}_{(2)}=-\frac{2\pi \sigma_cGa^2}{105 \chi^{7/2}} \Bigl\{ 6\chi^3(1+8k^4)
+ \chi^2 \bigl[ -16k^2(R^2+|z|^2)+a^2(9-26k^2-54k^4)  \bigr.  \Bigr. \\
\Bigl. \bigl. +3a|z|(2-16k^2-19k^4) \bigr] -a^2\chi \bigl[ 2R^2(2+11k^2+9k^4)+|z|^2(1-26k^2-27k^4) \bigr. \Bigr. \\
\Bigl. \bigl. -a|z|(13+82k^2+69k^4) -2a^2(7+28k^2+21k^4) \bigr] \Bigr. \\
\Bigl. -3a^3(a+|z|)(1+k^2)^2\bigl[ 2R^2+7(a+|z|)^2\bigr] \Bigr\} \mbox{,}
\end{multline}
where $\chi=R^2+(a+|z|)^2$ . Expressions for the circular velocity
$v^{(1,2)}_{c(2)}$ and the epicyclic frequency $\kappa^{(1,2)}_{(2)}$ follow from equation 
(\ref{eq_phi_r2}), and are listed in Appendix \ref{ap_D}.

\begin{figure}
\centering
\includegraphics[scale=0.85]{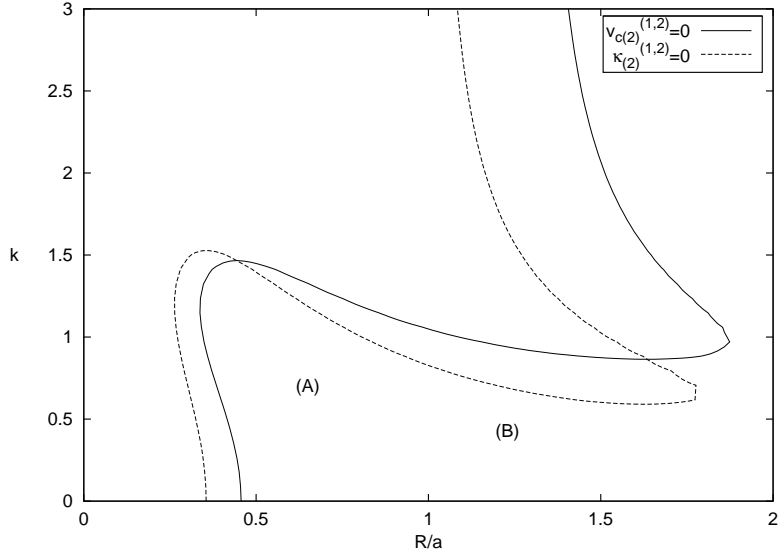}
\caption{The curves $v^{(1,2)}_{c(2)}=0$ (solid line) and $\kappa^{(1,2)}_{(2)}=0$
(dashed line) as functions of $k$ and $R/a$ for the first member of a family of double rings.} \label{fig_r5}
\end{figure}

\begin{figure}
\centering
\includegraphics[scale=0.85]{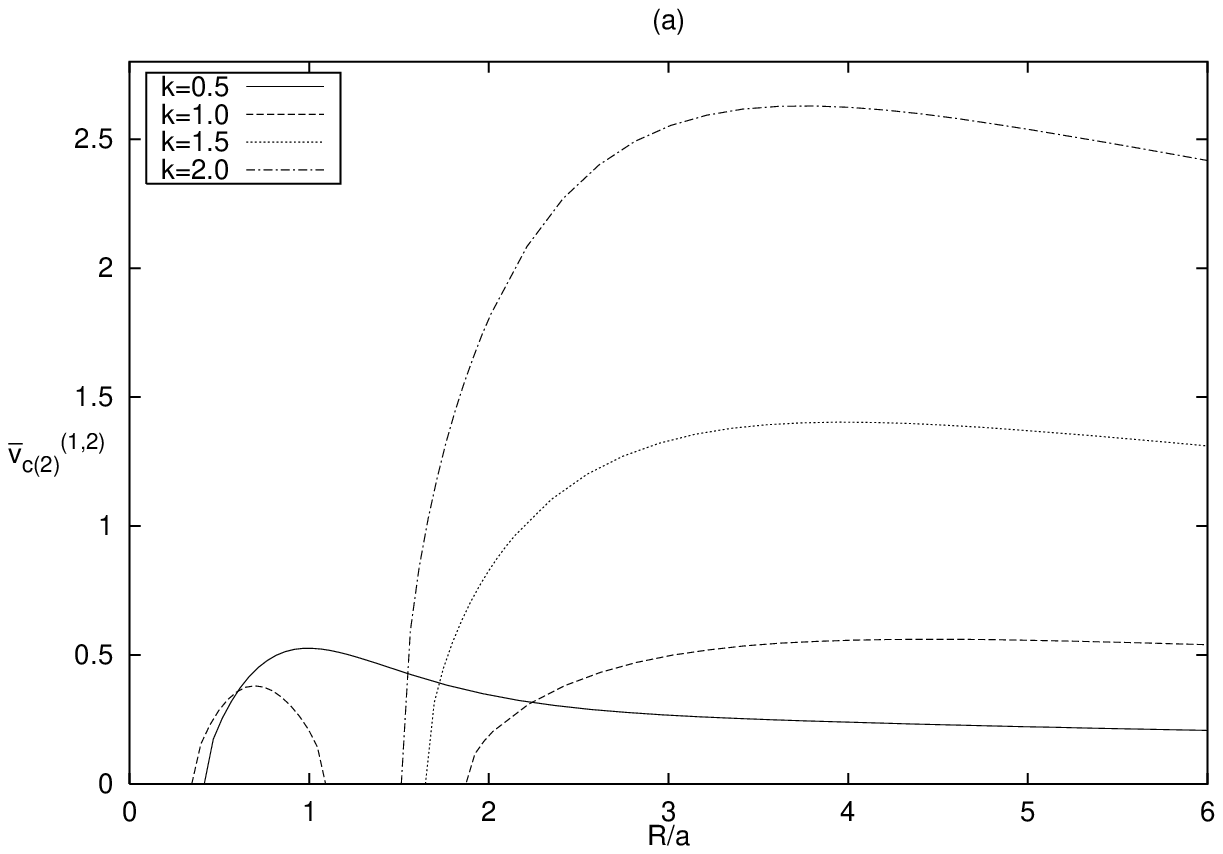}  \\
\vspace{0.1cm}
\includegraphics[scale=0.85]{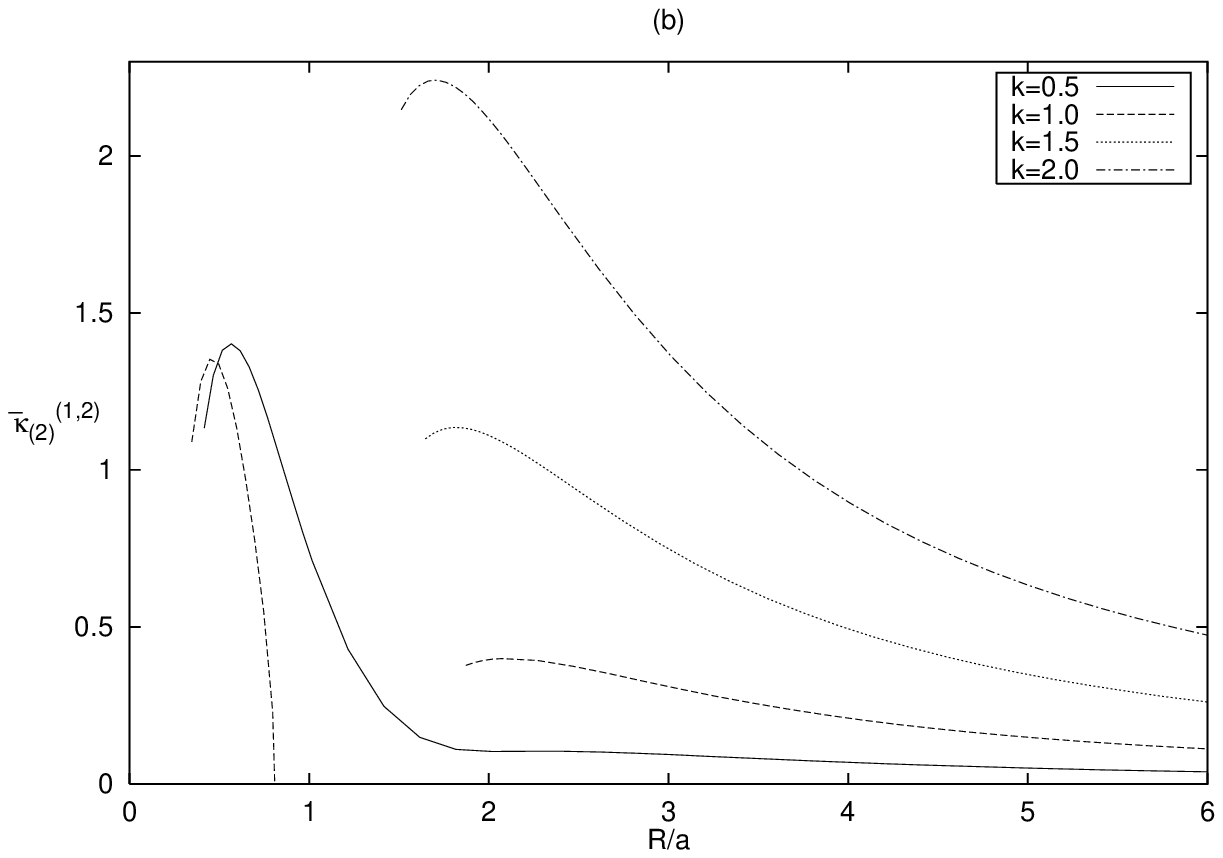}
\caption{(a) The circular velocity $\bar{v}^{(1,2)}_{c(2)}=v^{(1,2)}_{c(2)}/\sqrt{\sigma_cGa}$
equation (\ref{eq_vc_r2}) and (b) the epicyclic frequency
$\bar{\kappa}^{(1,2)}_{(2)}=\kappa^{(1,2)}_{(2)}/(\sigma_cG/a)^{1/2}$ equation
(\ref{eq_kappa_r2}) for the first member of a family of double rings 
as functions of $R/a$ for $k=0.5$, $k=1$, $k=1.5$ and $k=2$.} \label{fig_r6}
\end{figure}

The curves $v^{(1,2)}_{c(2)}=0$ (solid line) and $\kappa^{(1,2)}_{(2)}=0$
(dashed line) as functions of $k$ and $R/a$ are displayed in Fig.\ \ref{fig_r5}.
Circular orbits are possible in the region labeled (A) delimited by the solid curve and 
the axes; orbits are stable in the region (B) delimited by the dashed curve and the 
axes. Between $0.58 \lesssim k \lesssim 1.53$ there exist two regions where orbits are unstable 
and circular orbits are not possible: the first begins at the origin and the second is in an 
annular region. This is possibly due to the presence of the additional gap. 
We note that most possible circular orbits are 
also stable, except for the interval $0.86 \lesssim k \lesssim 1.47$ where some parts of the double rings
have unstable circular orbits.  Some curves of the dimensionless circular velocity 
$\bar{v}^{(1,2)}_{c(2)}=v^{(1,2)}_{c(2)}/\sqrt{\sigma_cGa}$ and dimensionless
epicyclic frequency $\bar{\kappa}^{(1,2)}_{(2)}=\kappa^{(1,2)}_{(2)}/(\sigma_cG/a)^{1/2}$
for $k=0.5$, $k=1$ $k=1.5$ and $k=2$ are depicted in Fig.\ \ref{fig_r6}(a) and 
Fig.\ \ref{fig_r6}(b), respectively. 

\subsection{Discs with Flat Rings} \label{ss_dr}

\begin{figure}
\centering
\includegraphics[scale=0.85]{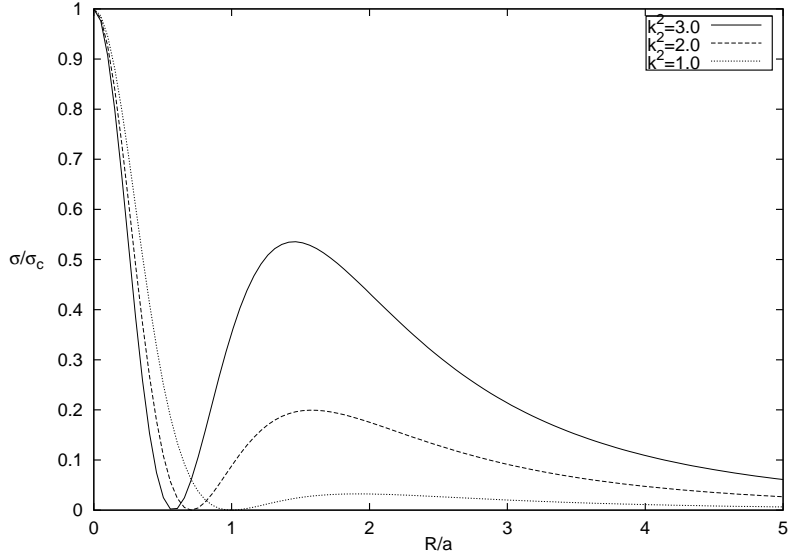}
\caption{The surface density $\sigma^{(2)}_{(d)}/\sigma_c$ for the first member of
a family of disc and ring (\ref{eq_sigma_dr}), as function of $R/a$, for $k^2=3$, $k^2=2$ and $k^2=1$.} \label{fig_r8}
\end{figure}

It is also possible to construct potential-density pairs for a Kuzmin-Toomre disc
surrounded by a concentric flat single ring. For this, we consider the superposition
\begin{equation} \label{eq_sigma_dr}
\sigma_{(d)}^{(n)}=\sigma_{n}-2k^2\sigma^{(1,n-1)}+k^4\sigma^{(2,n-2)}=
\frac{\sigma_c\left( 1-k^2R^2/a^2\right)^2}{\left(1+R^2/a^2 \right)^{n+3/2}} \mbox{.}
\end{equation}
Equation (\ref{eq_sigma_dr}) represents the density of a disc with radius $R/a=1/k$ and a flat ring
between $1/k<R/a<\infty$. The total mass of this superposition is
\begin{equation}
 \mathcal{M}^{(n)}_{(d)}=\frac{2\pi\sigma_ca^2}{(2n+1)(2n-1)(2n-3)}\bigl[ (2n-1)(2n-3)-
4k^2(2n-3)+8k^4 \bigr] \mbox{.}
\end{equation}
Again we study the first member of this family with $n=2$. 
In Fig.\ \ref{fig_r8} the non-dimensional surface density $\sigma^{(2)}_{(d)}/\sigma_c$
is shown for the values $k^2=3$, $k^2=2$ and $k^2=1$. For large values of $k$, the density of
the ring is more concentrated.
The associated potential $\phi^{(2)}_{(d)}$ can be expressed as
\begin{multline} \label{eq_phi_dr}
\phi^{(2)}_{(d)}=-\frac{2\pi \sigma_cGa^2}{15\chi^{5/2}} \Bigl\{
\chi^2(3-4k^2)+\chi \bigl[ a(a+|z|)(3-4k^2)+k^4(8R^2+8|z|^2+9a|z|)\bigr] \Bigr. \\
\Bigl. +a^2\bigl[ -R^2(1+2k^2)+(a+|z|)^2(2+4k^2+3k^4)\bigr] \Bigr\} \mbox{,}
\end{multline}
where $\chi=R^2+(a+|z|)^2$. By using equation (\ref{eq_phi_dr}), expressions for the circular velocity
$v^{(2)}_{c(d)}$ and the epicyclic frequency $\kappa^{(2)}_{(d)}$  can be calculated; they 
are given in Appendix \ref{ap_E}.

\begin{figure}
\centering
\includegraphics[scale=0.85]{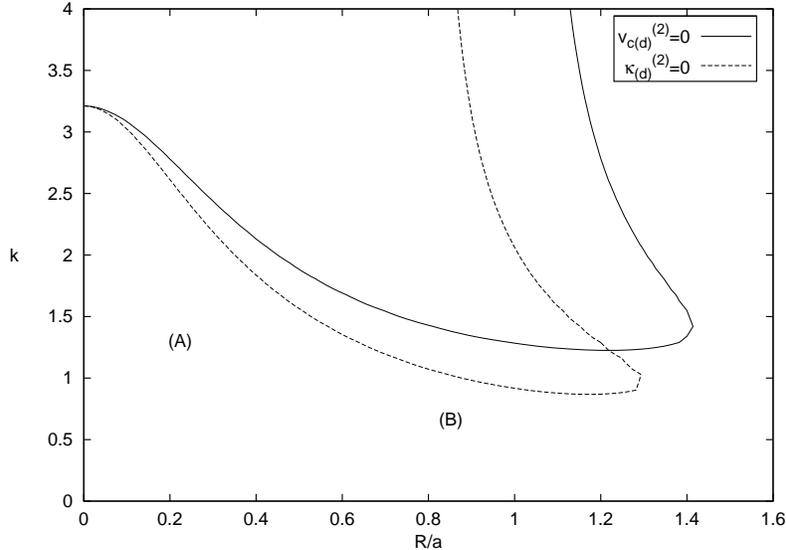}
\caption{The curves $v^{(2)}_{c(d)}=0$ (solid lines) and $\kappa^{(2)}_{(d)}=0$
(dashed lines) as functions of $k$ and $R/a$ for the first member of a superposition of
disc and ring.} \label{fig_r9}
\end{figure}

The curves $v^{(2)}_{c(d)}=0$ (solid line) and $\kappa^{(2)}_{(d)}=0$
(dashed line) as functions of $k$ and $R/a$ are shown in Fig.\ \ref{fig_r9}.
In the region (A), delimited by the solid curve and the axes, circular orbits are possible and 
in region (B), delimited by the dashed curve and the axes, we have stability. In the limit $k=0$, we have a pure disc 
that is stable and all circular orbits are possible. As the value of $k$ grows, the influence of 
the outer ring on the rotation curves and stability becomes more pronounced. In the interval $0 \leq k \lesssim 0.86$, all
circular orbits are also stable. When $1.22 \lesssim k \lesssim 3.21$, there exist possible circular 
orbits that are unstable. For $k \gtrsim 3.21$, all possible circular orbits are stable. Curves for the 
circular velocity and epicyclic frequency are qualitatively similar to those for the two-ring system. 
\subsection{A Planet and a Ring} \label{ss_rpl}

\begin{figure}
\centering
\includegraphics[scale=0.85]{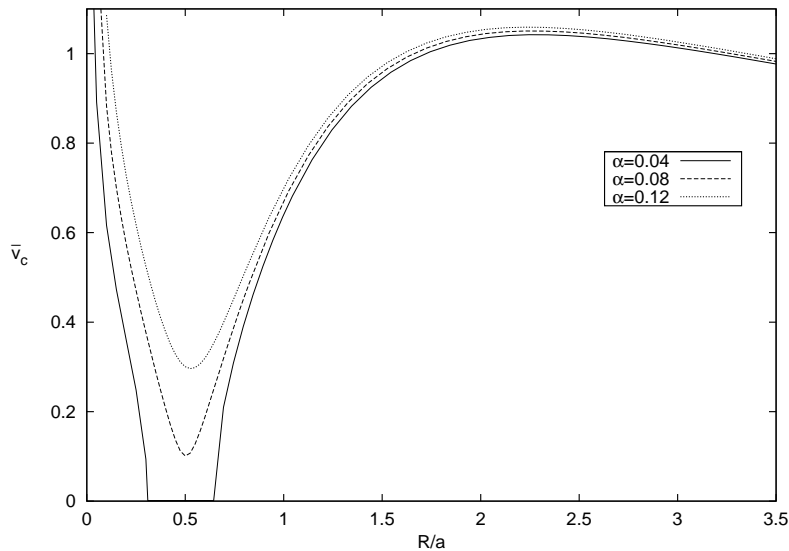}
\caption{The circular velocity $\bar{v}_c=v_c/\sqrt{\sigma_cGa}$
equation (\ref{eq_vc_rpl}) for a planet-ring system as function of $R/a$ for $\alpha=m/(\sigma_ca^2)=0.04$,
$\alpha=0.08$ and $\alpha=0.12$.} \label{fig_r12}
\end{figure}

Now we study a very simple example of a planet located at the centre of a ring. Let us 
consider a point mass with potential $\phi=-Gm/R$ and the first member of the family 
of single rings with potential $\phi^{(1,0)}$, equation (\ref{eq_r10}). The rotation curve  and
epicyclic frequency of particles orbiting this planet-ring system are given by
\begin{gather}
v_c= \left[ \frac{Gm}{R} + \frac{2\pi \sigma_cGa^2R^2(2R^2-a^2)}{3(R^2+a^2)^{5/2}} 
\right]^{1/2} \mbox{,} \label{eq_vc_rpl} \\
\kappa= \left[ \frac{Gm}{R^3}+\frac{2\pi \sigma_cGa^2}{3(R^2+a^2)^{7/2}}
(2R^4+13R^2a^2-4a^4) \right]^{1/2} \mbox{,} \label{eq_kappa_rpl} 
\end{gather}
The contribution of the planet to the circular velocity is always positive. As was pointed 
out in Sec.\ \ref{ss_one}, circular orbits were not possible near the ring's centre due
to a lack of mass in this region of space. But if the planet has enough mass, the square 
of the circular velocity can be made non-negative everywhere. This happens if 
$m/(\sigma_ca^2) \geq 4\pi \sqrt{5}/375 \approx 0.075$.
In Fig.\ \ref{fig_r12} we
display some rotation curves, where $\bar{v}_c=v_c/\sqrt{\sigma_cGa}$ is the dimensionless
velocity and where we defined $\alpha=m/(\sigma_ca^2)$.
By equation (\ref{eq_kappa_rpl}), the contribution of the central mass to the
epicyclic frequency is positive, so it helps
to stabilize the circular orbits in the plane of the ring. In fact, it can be shown that if
\begin{equation}
\frac{m}{\sigma_ca^2} \geq \frac{\pi}{3675}(65\sqrt{19}-41\sqrt{35})\sqrt{70-2\sqrt{665}}
\approx 0.15 \mbox{,}
\end{equation}
the epicyclic frequency is always non-negative. 
\section{Thick Rings} \label{sec_thick}

Until now all ring structures presented were flat, i.e., had infinitesimal thickness. 
One way to generate three-dimensional potential-density pairs from these flat ring models is
to employ the same transformation used by Miyamoto and Nagai \cite{m1} to ``inflate'' the Kuzmin-Toomre
discs. 

As a first example, we take the potential $\phi^{(1,0)}$ of the first member of the family of single rings
and apply a transformation $|z| \rightarrow \sqrt{z^2+b^2}$, where $b$ is a non-negative constant,
to equation (\ref{eq_r10}).
The resulting mass density is calculated directly from the Poisson equation
in cylindrical coordinates
\begin{equation}
\rho=\frac{1}{4\pi G} \left( \phi_{,RR}+\frac{\phi_{,R}}{R}+\phi_{,zz} \right) \mbox{.}
\end{equation}
We obtain
\begin{equation} \label{eq_rho10}
\rho^{(1,0)}=\frac{\sigma_ca^2b^2}{2\xi^3\bigl[ R^2+(a+\xi)^2 \bigr]^{7/2}} \Bigl\{ aR^4+R^2(a+\xi)
\left[ a^2+2\xi (3a+\xi) \right] +2\xi^2 (a+\xi )^3 \Bigr\} \mbox{,}
\end{equation}
where $\xi=\sqrt{z^2+b^2}$.
\begin{figure}
\centering
\includegraphics[scale=0.7]{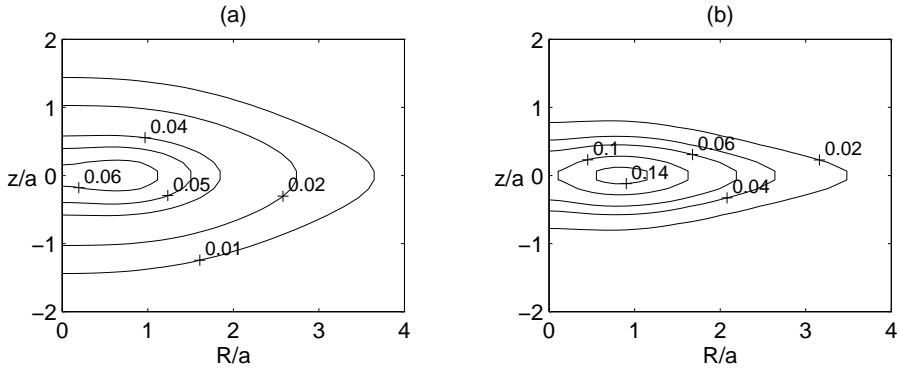}
\caption{Isodensity curves of the mass density $\rho^{(1,0)}/(\sigma_c/a)$ equation
(\ref{eq_rho10}) for a thick single ring as functions of $R/a$ and $z/a$. 
Parameters: (a) $b/a=1$ and (b) $b/a=0.5$.} \label{fig_r13}
\end{figure}

By inspection, the mass density equation (\ref{eq_rho10}) is non-negative and free of 
singularities. Fig.\ \ref{fig_r13}(a)--(b) show some isodensity curves of $\rho^{(1,0)}/(\sigma_c/a)$ 
as functions of $R/a$ and $z/a$ with parameters $b/a=1$ in Fig.\ \ref{fig_r13}(a) and 
$b/a=0.5$ in Fig.\ \ref{fig_r13}(b). For this last parameter value, it is seen that
we have a toroidal mass distribution. We also ``inflated'' the single ring with potential
$\phi^{(1,1)}$ and found a three-dimensional mass density with similar properties.

It is also interesting to compute the rotation curve and epicyclic frequency of particles 
orbiting the symmetry plane $z=0$ of the ``inflated'' ring. We get
\begin{gather} 
v_{c(\mathrm{th.})}^{(1,0)} = R \left\{ \frac{2\pi \sigma_cGa^2\left[ 2R^2-(a+b)(a-2b)\right]}
{3\left[ R^2+\left(a+b\right)^2\right]^{5/2}} \right\}^{1/2} \mbox{,} \label{eq_vc10_th}  \\
\kappa^{(1,0)}_{\mathrm{th.}}=\left\{ \frac{2\pi \sigma_cGa^2}{3\left[ R^2+(a+b)^2\right]^{7/2}}
\left[ 2R^4+R^2(a+b)(13a+10b)  \right. \right. \notag \\
\left. \left. -4(a+b)^3(a-2b)\right] \right\}^{1/2} \mbox{,} \label{eq_kappa10_th}
\end{gather}
where the subscript $\mathrm{th.}$ refers to thick structures. As the thickening parameter $b/a$ 
is increased, the regions with impossible and unstable circular orbits shrink until they disappear
when $b/a \geq 1/2$. On the other hand, Fig.\ \ref{fig_r13} suggest that prominent toroidal 
mass distributions are formed for $b/a \leq 1/2$, and these still have regions with impossible and unstable circular 
orbits.  

If we apply a Miyamoto-Nagai transformation to the potential of two concentric flat rings, we
expect to generate two toroidal mass distributions. For instance, the potential 
$\phi^{(1,2)}_{(2)}$ equation (\ref{eq_phi_r2}) yields a somewhat involved expression
for the density, 
\begin{multline} \label{eq_rho_r2}
\rho^{(1,2)}_{(2)}=\frac{\sigma_ca^2b^2}{70\xi^3 \bigl[ R^2+(a+\xi)^2\bigr]^{11/2}}
\Bigl\{ 35ak^4R^8+R^6\bigl[ 2\xi^3 (3-8k^2+24k^4) \bigr. \Bigr. \\ 
\Bigl. \bigl. +35ak^4\xi (11a+10\xi)+35k^2a^3(k^2-2) \bigr] 
+R^4(a+\xi) \bigl[ 6\xi^4 (3-8k^2+24k^4) \bigr. \Bigr. \\
+a\xi^3(48-128k^2+699k^4)+5a^2\xi^2 (7-112k^2+126k^4)
-35a^3\xi (1+20k^2)+35a^4(1-2k^2) \bigr] \Bigr. \\
\Bigl. +R^2(a+\xi)^3 \bigl[ 6\xi^4 (3-8k^2+24k^4)
+2a\xi^3(39-104k^2+172k^4)+5a^2\xi^2 (21-112k^2)  \bigr.   \Bigr. \\
\Bigl. \bigl. +5a^3(7a+56\xi) \bigr]+2\xi^2(a+\xi)^5\bigl[  \xi^2(3-8k^2+24k^4)
+2a\xi (9-24k^2+2k^4)+35a^2 \bigr] \Bigr\} \mbox{,}
\end{multline}
where $\xi=\sqrt{z^2+b^2}$.
\begin{figure}
\centering
\includegraphics[scale=0.7]{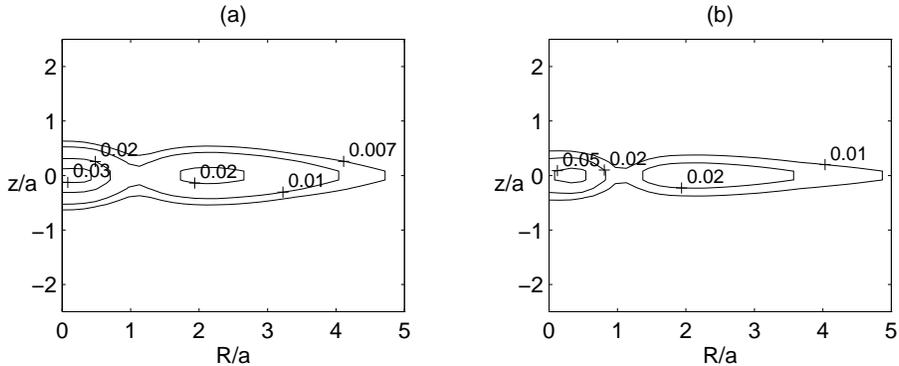}
\caption{Isodensity curves of the mass density $\rho^{(1,2)}_{(2)}/(\sigma_c/a)$ equation
(\ref{eq_rho_r2}) for a thick double ring as functions of $R/a$ and $z/a$. 
Parameters (a) $k=1$, $b/a=0.5$ and (b) $k=1$, $b/a=0.3$.} \label{fig_r14}
\end{figure}

Some isodensity curves of the mass density $\rho^{(1,2)}_{(2)}/(\sigma_c/a)$ are displayed 
in Fig.\ \ref{fig_r14}(a)--(b) with parameters $k=1$, 
$b/a=0.5$ in Fig.\ \ref{fig_r14}(a) and $k=1$, $b/a=0.3$ in Fig.\ \ref{fig_r14}(b). For this set
of parameters, the mass density is positive everywhere. We also tested it for some other
values of $k$ and $b/a$ and did not find regions with negative mass density or singularities.

\begin{figure}
\centering
\includegraphics[scale=0.8]{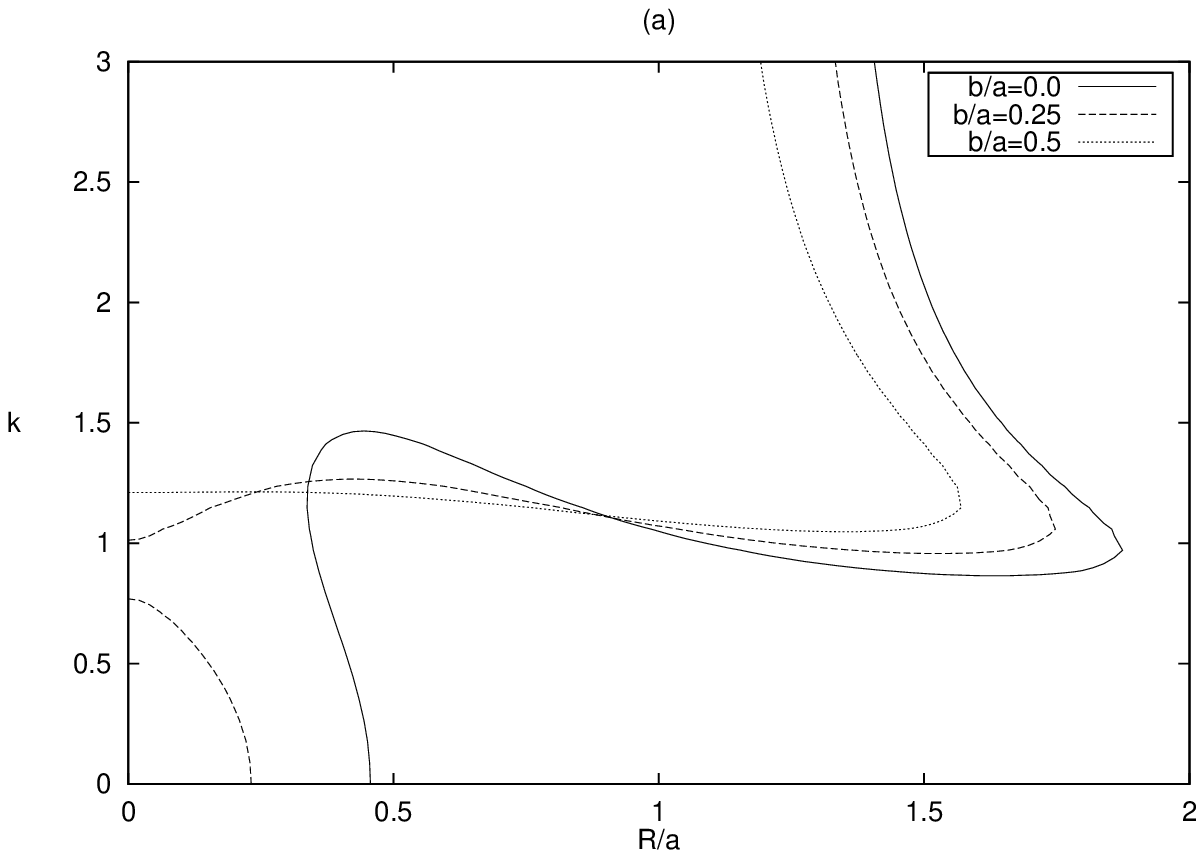}  \\
\vspace{0.1cm}
\includegraphics[scale=0.8]{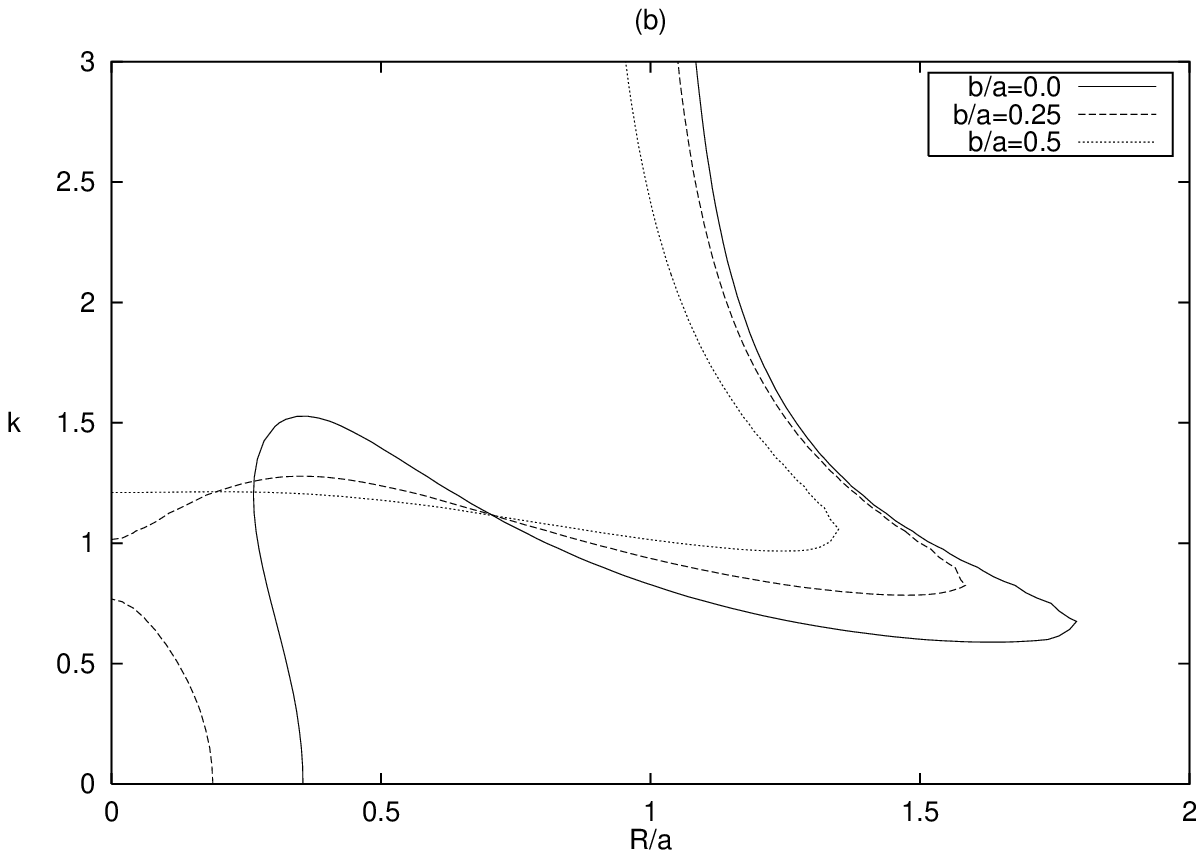}
\caption{(a) Curves of $v^{(1,2)}_{c(2)\mathrm{th.}}=0$ equation (\ref{eq_vc_2th}) and (b) 
$\kappa^{(1,2)}_{(2)\mathrm{th.}}=0$ equation
(\ref{eq_kappa_2th}) for a thick double ring as functions of $R/a$ and $k$ for $b/a=0$, $b/a=0.25$ and $b/a=0.5$.} \label{fig_r15}
\end{figure}
For this two-toroid system, we also computed the rotation curve and epicyclic frequency. The 
explicit expressions are given in Appendix \ref{ap_F}. Fig.\ \ref{fig_r15}(a) shows some 
zero velocity curves of equation (\ref{eq_vc_2th}) and Fig.\ \ref{fig_r15}(b) curves of zero epicyclic frequency 
equation (\ref{eq_kappa_2th}) as functions of $R/a$ and $k$ for the thickening parameters $b/a=0$, $b/a=0.25$ and $b/a=0.5$. 
Here, thickening also helps to stabilize circular orbits and shrinks the regions with 
negative square velocities.  

\section{Discussion} \label{sec_discuss}

We presented some families of analytical potential-density pairs that represent one or
two concentric flat rings that were constructed by superposing members of
the Kuzmin-Toomre family of discs, and also a disc surrounded by a flat ring.
All potential-density pairs can be expressed in terms of elementary functions.
Although the models have infinite extension, the surface density decays
with the distance very fast, and, in principle, one can insert a clear cut-off 
and consider them as finite. For these models, we studied the circular orbits of 
test particles concentric to the rings and made a first study of the stability
of these structures by radial perturbations of circular orbits (epicyclic frequency). 
In general the two-ring 
structures tend to be more unstable than the single rings, and a surrounding 
ring also helps to desestabilize a disc. On the other hand, if we put a planet located
at the centre of a ring it tends to increase stability in the radial direction.

We also presented two examples of ``inflated'' rings that were obtained by
applying a Miyamoto-Nagai like transformation on the above-mentioned single and
double flat rings. The resulting three-dimensional mass densities have a
toroidal shape, are non-negative and free of singularities. The
thickening of flat ring structures increases stability in the radial direction.

As was mentioned, a first stability test of the presented ring structures was based on
the study of stability of individual particles. The study of stability of orbits 
perpendicular to the plane of flat rings is not trivial because of the abrupt change of matter density as
a particle crosses the plane of the ring. For instance, Saa and Venegeroles \cite{s1} showed that the presence of a 
thin plane  destroys the integrability of orbits of particles that cross the plane and these orbits tend  to be chaotic. 
This effect is similar to the behaviour of the Chua circuit, a well-known case of chaotic system \cite{c3}. Recently,    
 \cite{c1} studied the stability of orbits crossing Morgan and Morgan discs, thin finite dics.   

A more realistic approach should 
take into account collective effects. The stability of the
gravitating system will be determined by the behaviour of the  perturbation of the  
matter content of  the disk. Some examples of a General Relativistic stability analysis of discs
can be found in \cite{u1,u2}.

\section*{Acknowledgments}

We thank FAPESP for financial support, PSL also thanks CNPq. This research has 
made use of SAO/NASA's Astrophysics Data System abstract service, which is gratefully 
acknowledged.

\appendix 
\section{Potential-density pairs for Kuzmin-Toomre discs} \label{ap_A}

The explicit expressions for the density and potentials for Kuzmin-Toomre 
discs with with $n=1,2,3$ are
\begin{align}
\sigma_{1} &=\frac{3a^3M}{2\pi (R^2+a^2)^{5/2}} \text{,} \qquad
\phi_{1}=-\frac{GM}{\bigl[ R^2+(a+|z|)^2\bigr]^{3/2}}\left( R^2+|z|^2+2a^2+3a|z|
\right) \mbox{,} \\
\sigma_{2} &=\frac{5a^5M}{2\pi (R^2+a^2)^{7/2}} \text{,} \qquad
\phi_{2}=-\frac{GM}{3\bigl[ R^2+(a+|z|)^2\bigr]^{5/2}}\Bigl\{ 3\bigl[ R^2+(a+|z|)^2
\bigr] \Bigr. \notag \\
& \Bigl. \times \bigl[ R^2+(a+|z|)^2+a(a+|z|)\bigr]-a^2\bigl[ R^2-2(a+|z|)^2\bigr] \Bigr\} \mbox{,} \\
\sigma_{3} &=\frac{7a^7M}{2\pi (R^2+a^2)^{9/2}} \text{,} \qquad
\phi_{3}=-\frac{GM}{5\bigl[ R^2+(a+|z|)^2\bigr]^{7/2}} \Bigl\{ 5\bigl[ R^2+(a+|z|)^2
\bigr]^3 \Bigr. \notag \\
& \Bigl. +a\bigl[ R^2+(a+|z|)^2 \bigr]^2(4a+5|z|)+a^2\bigl[ R^2+(a+|z|)^2 \bigr]
\bigl[ -R^2+5|z|(a+|z|)\bigr]  \Bigr. \notag \\
& \Bigl. +a^3(a+|z|)\bigl[ 2R^2+7(a+|z|)^2 \bigr] \Bigr\} \mbox{.}
\end{align}
\section{Potential-density pairs for flat one rings} \label{ap_B}

The explicit expressions for the density and potentials of four members of flat 
one rings with $m=1,2$; $n=0,1$ read 
\begin{align}
\sigma^{(1,0)} &=\frac{\sigma_c\left( R/a \right)^2}{\left(1+R^2/a^2 \right)^{5/2}} \text{,} \qquad
\phi^{(1,0)}=-\frac{2\pi \sigma_cGa^2}{3\bigl[ R^2+(a+|z|)^2\bigr]^{3/2}} \left( 2R^2+2|z|^2 \right. \notag \\
& \left. +a^2+3a|z| \right) \mbox{,} \label{eq_r10} \\
\sigma^{(1,1)} &=\frac{\sigma_c\left( R/a \right)^2}{\left(1+R^2/a^2 \right)^{7/2}} \text{,} \qquad
\phi^{(1,1)}=-\frac{2\pi \sigma_cGa^2}{15\bigl[ R^2+(a+|z|)^2\bigr]^{5/2}} \Bigl[ R^2(2R^2+2|z|^2 \Bigr. \notag \\
& \Bigl. +5a^2+6a|z|)+2(a+|z|)^2(R^2+|z|^2+a^2+3a|z|) \Bigr] \mbox{,} \label{eq_r11} \\
\sigma^{(2,0)} &=\frac{\sigma_c\left( R/a \right)^4}{\left(1+R^2/a^2 \right)^{7/2}} \text{,} \qquad
\phi^{(2,0)}=-\frac{2\pi \sigma_cGa^2}{15\bigl[ R^2+(a+|z|)^2\bigr]^{5/2}} \Bigl\{
\bigl[ R^2+(a+|z|)^2\bigr] \Bigr. \notag \\
& \Bigl. \times (8R^2+8|z|^2+9a|z|)+3a^2(a+|z|)^2 \Bigr\} \mbox{,} \label{eq_r20} \\
\sigma^{(2,1)} &=\frac{\sigma_c\left( R/a \right)^4}{\left(1+R^2/a^2 \right)^{9/2}} \text{,} \qquad
\phi^{(2,1)}=-\frac{2\pi \sigma_cGa^2}{105\bigl[ R^2+(a+|z|)^2\bigr]^{7/2}} \Bigl\{
\bigl[ R^2+(a+|z|)^2\bigr]^2  \Bigr. \notag \\
& \Bigl. \times (8R^2+8|z|^2+13a^2+24a|z|) 
+a^2\bigl[ R^2+(a+|z|)^2\bigr] \Bigr. \notag \\
& \Bigl. \times (11R^2-13|z|^2-28a^2-41a|z|)+3a^3(a+|z|)\bigl[ 2R^2+7(a+|z|)^2\bigr] \Bigr\} \mbox{.} \label{eq_r21}
\end{align}
\section{Rotation curves and epicyclic frequency for flat one rings} \label{ap_C}

The expressions for the circular velocity $v_c^{(m,n)}$ and the regions with non-negative 
velocity for the members of flat one rings (\ref{eq_r10})--(\ref{eq_r21}) are given by
\begin{gather}
v_c^{(1,0)} =R \left[ \frac{2\pi \sigma_cGa^2(2R^2-a^2)}{3(R^2+a^2)^{5/2}} \right]^{1/2} 
\text{, } v_c^{(1,0)} \geq 0 \text{ for } \frac{R}{a} \geq \frac{\sqrt{2}}{2} \approx 0.71 \mbox{,} \label{eq_vc10} \\
v_c^{(1,1)} =R \left[ \frac{2\pi \sigma_cGa^2}{15(R^2+a^2)^{7/2}}
(2R^4+13R^2a^2-4a^4)\right]^{1/2} \mbox{,} \notag \\
 v_c^{(1,1)} \geq 0 \text{ for }
\frac{R}{a} \geq \frac{\sqrt{\sqrt{201}-13}}{2} \approx 0.54 \mbox{,} \\
v_c^{(2,0)} =R \left[ \frac{2\pi \sigma_cGa^2}{15(R^2+a^2)^{7/2}}
(8R^4-8R^2a^2-a^4)\right]^{1/2} \mbox{,} \notag \\
v_c^{(2,0)} \geq 0 \text{ for }
\frac{R}{a} \geq \frac{\sqrt{\sqrt{6}+2}}{2} \approx 1.05 \mbox{,} \\
v_c^{(2,1)} =R \left[ \frac{2\pi \sigma_cGa^2}{105(R^2+a^2)^{9/2}}
(8R^6+72R^4a^2-45R^2a^4-4a^6)\right]^{1/2} \mbox{,} \notag \\
v_c^{(2,1)} \geq 0 \text{ for } \frac{R}{a}  \gtrsim 0.81 \mbox{.} \label{eq_vc21}
\end{gather}
The corresponding expressions for the epicyclic frequency $\kappa^{(m,n)}$ and the 
regions of stability read 
\begin{gather}
\kappa^{(1,0)}=\left[ \frac{2\pi \sigma_cGa^2}{3(R^2+a^2)^{7/2}}
(2R^4+13R^2a^2-4a^4) \right]^{1/2} \mbox{,} \notag \\
\text{stable for } \frac{R}{a}
\geq  \frac{\sqrt{\sqrt{201}-13}}{2} \approx 0.54 \mbox{,} \label{eq_kappa10} \\
\kappa^{(1,1)} =\left[ \frac{2\pi \sigma_cGa^2}{15(R^2+a^2)^{9/2}}
(2R^6+3R^4a^2+90R^2a^4-16a^6) \right]^{1/2}  \mbox{,} \notag \\
\text{stable for } \frac{R}{a} \gtrsim 0.42 \mbox{,} \\
\kappa^{(2,0)} =\left[ \frac{2\pi \sigma_cGa^2}{15(R^2+a^2)^{9/2}}
(8R^6+72R^4a^2-45R^2a^4-4a^6) \right]^{1/2} \mbox{,} \notag \\
\text{stable for } \frac{R}{a} \gtrsim 0.81 \mbox{,} \\
\kappa^{(2,1)} =\left[ \frac{2\pi \sigma_cGa^2}{105(R^2+a^2)^{11/2}}
(8R^8+8R^6a^2+711R^4a^4-250R^2a^6-16a^8) \right]^{1/2} \mbox{,} \notag \\
\text{stable for } \frac{R}{a} \gtrsim 0.64 \mbox{.} \label{eq_kappa21}
\end{gather}
\section{Rotation curves and epicyclic frequency for flat two rings} \label{ap_D}

The expressions for the circular velocity $v^{(1,2)}_{c(2)}$ and the epicyclic frequency $\kappa^{(1,2)}_{(2)}$
for a particular member of two flat rings studied in Section \ref{ss_two} are
given by
\begin{gather}
v^{(1,2)}_{c(2)}=R \left\{ \frac{2\pi \sigma_cGa^2}{105(R^2+a^2)^{9/2}} \bigl[
2R^6(3-8k^2+24k^4)-3R^4a^2(-11+48k^2+24k^4) \bigr. \right. \notag \\
\left. \bigl. +18R^2a^4(6+5k^2-k^4)-a^6(24-8k^2+3k^4) \bigr] \right\}^{1/2} \mbox{,} \label{eq_vc_r2} \\
\kappa^{(1,2)}_{(2)} = \left\{ \frac{2\pi \sigma_cGa^2}{105(R^2+a^2)^{11/2}} \bigl[
2R^8(3-8k^2+24k^4)+R^6a^2(27-16k^2+552k^4) \bigr. \right. \notag \\
\left. \bigl. -6R^4a^4(10+237k^2+87k^4)+R^2a^6(768+500k^2-93k^4) \bigr. \right. \notag \\
\left. \bigl. -4a^8(24-8k^2+3k^4) \bigr] \right\}^{1/2} \mbox{.} \label{eq_kappa_r2} 
\end{gather}
\section{Rotation curves and epicyclic frequency for discs with flat rings} \label{ap_E}

The expressions for the circular velocity $v^{(2)}_{c(d)}$ and the epicyclic frequency $\kappa^{(2)}_{(d)}$
for a particular member of the disc with a flat ring studied in Section \ref{ss_dr} are
given by
\begin{gather}
v^{(2)}_{c(d)}=R \left\{ \frac{2\pi \sigma_cGa^2}{15(R^2+a^2)^{7/2}} \bigl[
R^4(3-4k^2+8k^4)-2R^2a^2(-6+13k^2+4k^4) \bigr. \right. \notag \\
\left. \bigl. +a^4(24+8k^2-k^4) \bigr] \right\}^{1/2} \mbox{,} \label{eq_vc_dr} \\
\kappa^{(2)}_{(d)} = \left\{ \frac{2\pi \sigma_cGa^2}{15(R^2+a^2)^{9/2}} \bigl[
R^6(3-4k^2+8k^4)+6R^4a^2(2-k^2+12k^4) \bigr. \right. \notag \\
\left. \bigl. -45R^2a^4k^2(4+k^2)+4a^6(24+8k^2-k^4) \bigr] \right\}^{1/2} \mbox{.} \label{eq_kappa_dr} 
\end{gather}
\section{Rotation curves and epicyclic frequency for thick two rings} \label{ap_F}

The expressions for the circular velocity $v^{(1,2)}_{c(2)\mathrm{th.}}$ and the epicyclic 
frequency $\kappa^{(1,2)}_{(2)\mathrm{th.}}$
for a particular member of thick two rings studied in Section \ref{sec_thick} are
given by
\begin{gather}
v^{(1,2)}_{c(2)\mathrm{th.}}=R \left\{ \frac{2\pi \sigma_cGa^2}{105\left[ R^2+(a+b)^2\right]^{9/2}} \bigl\{
2R^6(3-8k^2+24k^4)  \bigr. \right. \notag \\ 
\left. \bigl. +3R^4\bigl[ -a^2(-11+48k^2+24k^4)+3ab(6-16k^2+13k^4)
 +2b^2(3-8k^2+24k^4) \bigr] \bigr. \right. \notag \\ 
\left. \bigl. +3R^2(a+b)\bigl[ 6a^3(6+5k^2-k^4)+2b^3(3-8k^2+24k^4)+10ab^2(3-8k^2+3k^4)
 \bigr. \bigr. \right. \notag \\
\left. \bigl. \bigl. -a^2b(-45+64k^2+39k^4)\bigr]
-(a+b)^3 \bigl[ a^3(24-8k^2+3k^4)-2b^3(3-8k^2+24k^4) \bigr. \bigr. \right. \notag \\
\left. \bigl. \bigl. +3ab^2(-12+32k^2+9k^4)
-6a^2b(11+8k^2-3k^4) \bigr] \bigr\} \right\}^{1/2} \mbox{,} \label{eq_vc_2th} \\
\kappa^{(1,2)}_{(2)\mathrm{th.}} = \left\{ \frac{2\pi \sigma_cGa^2}{105\left[ R^2+(a+b)^2\right]^{11/2}} \bigl\{
2R^8(3-8k^2+24k^4) \bigr. \right. \notag \\
\left. \bigl.  +R^6 \bigl[ a^2(27-16k^2+552k^4)+ab(66-176k^2+843k^4) 
+14b^2(3-8k^2+24k^4) \bigr] \bigr. \right. \notag \\
\left. \bigl.  +3R^4(a+b) \bigl[ -2a^3(10+237k^2+87k^4)+a^2b(97-576k^2+237k^4)
\bigr. \bigr. \right. \notag \\
\left. \bigl. \bigl. +2ab^2(51-136k^2+303k^4) 
+10b^3(3-8k^2+24k^4) \bigr] +R^2(a+b)^3 \bigr. \right. \notag \\
\left. \bigl. \times \bigl[ a^3(768+500k^2-93k^4)-12a^2b(-40+116k^2+51k^4)
 \bigr. \bigr. \right. \notag \\
\left. \bigl. \bigl. +15ab^2(24-64k^2+45k^4)+26b^3(3-8k^2+24k^4) \bigr]-4(a+b)^5  \bigr. \right. \notag \\
\left. \bigl. \times \bigl[ a^3(24-8k^2+3k^4)+6a^2b(-11-8k^2+3k^4)+3ab^2(-12+32k^2+9k^4) \bigr. \bigr. \right. \notag \\
\left. \bigl. \bigl. -2b^3(3-8k^2+24k^4) \bigr] \bigr\} \right\}^{1/2} \mbox{.} \label{eq_kappa_2th} 
\end{gather}
\end{document}